\documentstyle[11pt]{article}
\newcommand{\be}{\begin{equation}}
\newcommand{\dis}{\displaystyle}
\newcommand{\ee}{\end{equation}}
\newcommand{\bea}{\begin{eqnarray}}
\newcommand{\eea}{\end{eqnarray}}
\newcommand{\nn}{\nonumber \\}
\newcommand{\vv}{{\scriptstyle \cal V}}
\newcommand{\ag}{{\scriptstyle \cal G}}
\newcommand{\om}{\omega}
\newcommand{\cc}{{(\tilde{c})}}
\newcommand{\ccc}{{(c)}}
\newcommand{\bbx}{\stackrel{\scriptstyle {}^\land_{} }{\Box}}

\immediate\write16{<WARNING: FEYNMAN macros work only with emTeX-dvivers
                    (dviscr.exe, dvihplj.exe, dvidot.exe, etc.) >}
\newdimen\Lengthunit
\newcount\Nhalfperiods
\catcode`\*=11
\newdimen\L*   \newdimen\d*   \newdimen\d**
\newdimen\dm*  \newdimen\dd*  \newdimen\dt*
\newdimen\a*   \newdimen\b*   \newdimen\c*
\newdimen\a**  \newdimen\b**
\newdimen\xL*  \newdimen\yL*
\newcount\k*   \newcount\l*   \newcount\m*
\newcount\n*   \newcount\dn*  \newcount\r*
\newcount\N*   \newcount\*one \newcount\*two  \*one=1 \*two=2
\newcount\*ths \*ths=1000
\def\GRAPH(hsize=#1)#2{\hbox to #1\Lengthunit{#2\hss}}
\def\Linewidth#1{\special{em:linewidth #1}}
\Linewidth{.4pt}
\def\sm*{\special{em:moveto}}
\def\sl*{\special{em:lineto}}
\newbox\spm*   \newbox\spl*
\setbox\spm*\hbox{\sm*}
\setbox\spl*\hbox{\sl*}
\def\mov#1(#2,#3)#4{\rlap{\L*=#1\Lengthunit\kern#2\L*\raise#3\L*\hbox{#4}}}
\def\smov#1(#2,#3)#4{\rlap{\L*=#1\Lengthunit
\xL*=\xscale\L*\yL*=\yscale\L*\kern#2\xL*\raise#3\yL*\hbox{#4}}}
\def\mov*(#1,#2)#3{\rlap{\kern#1\raise#2\hbox{#3}}}
\def\lin#1(#2,#3){\rlap{\sm*\mov#1(#2,#3){\sl*}}}
\def\arr*(#1,#2,#3){\mov*(#1\dd*,#1\dt*){%
\sm*\mov*(#2\dd*,#2\dt*){\mov*(#3\dt*,-#3\dd*){\sl*}}%
\sm*\mov*(#2\dd*,#2\dt*){\mov*(-#3\dt*,#3\dd*){\sl*}}}}
\def\arrow#1(#2,#3){\rlap{\lin#1(#2,#3)\mov#1(#2,#3){%
\d**=-.012\Lengthunit\dd*=#2\d**\dt*=#3\d**%
\arr*(1,10,4)\arr*(3,8,4)\arr*(4.8,4.2,3)}}}
\def\arrlin#1(#2,#3){\rlap{\L*=#1\Lengthunit\L*=.5\L*%
\lin#1(#2,#3)\mov*(#2\L*,#3\L*){\arrow.1(#2,#3)}}}
\def\dasharrow#1(#2,#3){\rlap{%
{\Lengthunit=0.9\Lengthunit\dashlin#1(#2,#3)\mov#1(#2,#3){\sm*}}%
\mov#1(#2,#3){\sl*\d**=-.012\Lengthunit\dd*=#2\d**\dt*=#3\d**%
\arr*(1,10,4)\arr*(3,8,4)\arr*(4.8,4.2,3)}}}
\def\clap#1{\hbox to 0pt{\hss #1\hss}}
\def\ind(#1,#2)#3{\rlap{%
\d*=.1\Lengthunit\kern#1\d*\raise#2\d*\hbox{\lower2pt\clap{$#3$}}}}
\def\sh*(#1,#2)#3{\rlap{%
\dm*=\the\n*\d**\xL*=\xscale\dm*\yL*=\yscale\dm*
\kern#1\xL*\raise#2\yL*\hbox{#3}}}
\def\calcnum*#1(#2,#3){\a*=1000sp\b*=1000sp\a*=#2\a*\b*=#3\b*%
\ifdim\a*<0pt\a*-\a*\fi\ifdim\b*<0pt\b*-\b*\fi%
\ifdim\a*>\b*\c*=.96\a*\advance\c*.4\b*%
\else\c*=.96\b*\advance\c*.4\a*\fi%
\k*\a*\multiply\k*\k*\l*\b*\multiply\l*\l*%
\m*\k*\advance\m*\l*\n*\c*\r*\n*\multiply\n*\n*%
\dn*\m*\advance\dn*-\n*\divide\dn*2\divide\dn*\r*%
\advance\r*\dn*%
\c*=\the\Nhalfperiods5sp\c*=#1\c*\ifdim\c*<0pt\c*-\c*\fi%
\multiply\c*\r*\N*\c*\divide\N*10000}
\def\dashlin#1(#2,#3){\rlap{\calcnum*#1(#2,#3)%
\d**=#1\Lengthunit\ifdim\d**<0pt\d**-\d**\fi%
\divide\N*2\multiply\N*2\advance\N*1%
\divide\d**\N*\sm*\n*\*one\sh*(#2,#3){\sl*}%
\loop\advance\n*\*one\sh*(#2,#3){\sm*}\advance\n*\*one\sh*(#2,#3){\sl*}%
\ifnum\n*<\N*\repeat}}
\def\dashdotlin#1(#2,#3){\rlap{\calcnum*#1(#2,#3)%
\d**=#1\Lengthunit\ifdim\d**<0pt\d**-\d**\fi%
\divide\N*2\multiply\N*2\advance\N*1\multiply\N*2%
\divide\d**\N*\sm*\n*\*two\sh*(#2,#3){\sl*}\loop%
\advance\n*\*one\sh*(#2,#3){\kern-1.48pt\lower.5pt\hbox{\rm.}}%
\advance\n*\*one\sh*(#2,#3){\sm*}%
\advance\n*\*two\sh*(#2,#3){\sl*}\ifnum\n*<\N*\repeat}}
\def\shl*(#1,#2)#3{\kern#1#3\lower#2#3\hbox{\unhcopy\spl*}}
\def\trianglin#1(#2,#3){\rlap{\toks0={#2}\toks1={#3}\calcnum*#1(#2,#3)%
\dd*=.57\Lengthunit\dd*=#1\dd*\divide\dd*\N*%
\d**=#1\Lengthunit\ifdim\d**<0pt\d**-\d**\fi%
\multiply\N*2\divide\d**\N*\advance\N*-1\sm*\n*\*one\loop%
\shl**{\dd*}\dd*-\dd*\advance\n*2%
\ifnum\n*<\N*\repeat\n*\N*\advance\n*1\shl**{0pt}}}
\def\wavelin#1(#2,#3){\rlap{\toks0={#2}\toks1={#3}\calcnum*#1(#2,#3)%
\dd*=.23\Lengthunit\dd*=#1\dd*\divide\dd*\N*%
\d**=#1\Lengthunit\ifdim\d**<0pt\d**-\d**\fi%
\multiply\N*4\divide\d**\N*\sm*\n*\*one\loop%
\shl**{\dd*}\dt*=1.3\dd*\advance\n*1%
\shl**{\dt*}\advance\n*\*one%
\shl**{\dd*}\advance\n*\*two%
\dd*-\dd*\ifnum\n*<\N*\repeat\n*\N*\shl**{0pt}}}
\def\w*lin(#1,#2){\rlap{\toks0={#1}\toks1={#2}\d**=\Lengthunit\dd*=-.12\d**%
\N*8\divide\d**\N*\sm*\n*\*one\loop%
\shl**{\dd*}\dt*=1.3\dd*\advance\n*\*one%
\shl**{\dt*}\advance\n*\*one%
\shl**{\dd*}\advance\n*\*one%
\shl**{0pt}\dd*-\dd*\advance\n*1\ifnum\n*<\N*\repeat}}
\def\l*arc(#1,#2)[#3][#4]{\rlap{\toks0={#1}\toks1={#2}\d**=\Lengthunit%
\dd*=#3.037\d**\dd*=#4\dd*\dt*=#3.049\d**\dt*=#4\dt*\ifdim\d**>16mm%
\d**=.25\d**\n*\*one\shl**{-\dd*}\n*\*two\shl**{-\dt*}\n*3\relax%
\shl**{-\dd*}\n*4\relax\shl**{0pt}\else\ifdim\d**>5mm%
\d**=.5\d**\n*\*one\shl**{-\dt*}\n*\*two\shl**{0pt}%
\else\n*\*one\shl**{0pt}\fi\fi}}
\def\d*arc(#1,#2)[#3][#4]{\rlap{\toks0={#1}\toks1={#2}\d**=\Lengthunit%
\dd*=#3.037\d**\dd*=#4\dd*\d**=.25\d**\sm*\n*\*one\shl**{-\dd*}%
\n*3\relax\sh*(#1,#2){\xL*=\xscale\dd*\yL*=\yscale\dd*
\kern#2\xL*\lower#1\yL*\hbox{\sm*}}%
\n*4\relax\shl**{0pt}}}
\def\arc#1[#2][#3]{\rlap{\Lengthunit=#1\Lengthunit%
\sm*\l*arc(#2.1914,#3.0381)[#2][#3]%
\smov(#2.1914,#3.0381){\l*arc(#2.1622,#3.1084)[#2][#3]}%
\smov(#2.3536,#3.1465){\l*arc(#2.1084,#3.1622)[#2][#3]}%
\smov(#2.4619,#3.3086){\l*arc(#2.0381,#3.1914)[#2][#3]}}}
\def\dasharc#1[#2][#3]{\rlap{\Lengthunit=#1\Lengthunit%
\d*arc(#2.1914,#3.0381)[#2][#3]%
\smov(#2.1914,#3.0381){\d*arc(#2.1622,#3.1084)[#2][#3]}%
\smov(#2.3536,#3.1465){\d*arc(#2.1084,#3.1622)[#2][#3]}%
\smov(#2.4619,#3.3086){\d*arc(#2.0381,#3.1914)[#2][#3]}}}
\def\wavearc#1[#2][#3]{\rlap{\Lengthunit=#1\Lengthunit%
\w*lin(#2.1914,#3.0381)%
\smov(#2.1914,#3.0381){\w*lin(#2.1622,#3.1084)}%
\smov(#2.3536,#3.1465){\w*lin(#2.1084,#3.1622)}%
\smov(#2.4619,#3.3086){\w*lin(#2.0381,#3.1914)}}}
\def\shl**#1{\c*=\the\n*\d**\d*=#1%
\a*=\the\toks0\c*\b*=\the\toks1\d*\advance\a*-\b*%
\b*=\the\toks1\c*\d*=\the\toks0\d*\advance\b*\d*%
\a*=\xscale\a*\b*=\yscale\b*%
\raise\b*\rlap{\kern\a*\unhcopy\spl*}}
\def\wlin*#1(#2,#3)[#4]{\rlap{\toks0={#2}\toks1={#3}%
\c*=#1\l*\c*\c*=.01\Lengthunit\m*\c*\divide\l*\m*%
\c*=\the\Nhalfperiods5sp\multiply\c*\l*\N*\c*\divide\N*\*ths%
\divide\N*2\multiply\N*2\advance\N*1%
\dd*=.002\Lengthunit\dd*=#4\dd*\multiply\dd*\l*\divide\dd*\N*%
\d**=#1\multiply\N*4\divide\d**\N*\sm*\n*\*one\loop%
\shl**{\dd*}\dt*=1.3\dd*\advance\n*\*one%
\shl**{\dt*}\advance\n*\*one%
\shl**{\dd*}\advance\n*\*two%
\dd*-\dd*\ifnum\n*<\N*\repeat\n*\N*\shl**{0pt}}}
\def\wavebox#1{\setbox0\hbox{#1}%
\a*=\wd0\advance\a*14pt\b*=\ht0\advance\b*\dp0\advance\b*14pt%
\hbox{\kern9pt%
\mov*(0pt,\ht0){\mov*(-7pt,7pt){\wlin*\a*(1,0)[+]\wlin*\b*(0,-1)[-]}}%
\mov*(\wd0,-\dp0){\mov*(7pt,-7pt){\wlin*\a*(-1,0)[+]\wlin*\b*(0,1)[-]}}%
\box0\kern9pt}}
\def\rectangle#1(#2,#3){%
\lin#1(#2,0)\lin#1(0,#3)\mov#1(0,#3){\lin#1(#2,0)}\mov#1(#2,0){\lin#1(0,#3)}}
\def\dashrectangle#1(#2,#3){\dashlin#1(#2,0)\dashlin#1(0,#3)%
\mov#1(0,#3){\dashlin#1(#2,0)}\mov#1(#2,0){\dashlin#1(0,#3)}}
\def\waverectangle#1(#2,#3){\L*=#1\Lengthunit\a*=#2\L*\b*=#3\L*%
\ifdim\a*<0pt\a*-\a*\def\x*{-1}\else\def\x*{1}\fi%
\ifdim\b*<0pt\b*-\b*\def\y*{-1}\else\def\y*{1}\fi%
\wlin*\a*(\x*,0)[-]\wlin*\b*(0,\y*)[+]%
\mov#1(0,#3){\wlin*\a*(\x*,0)[+]}\mov#1(#2,0){\wlin*\b*(0,\y*)[-]}}
\def\calcparab*{%
\ifnum\n*>\m*\k*\N*\advance\k*-\n*\else\k*\n*\fi%
\a*=\the\k* sp\a*=10\a*\b*\dm*\advance\b*-\a*\k*\b*%
\a*=\the\*ths\b*\divide\a*\l*\multiply\a*\k*%
\divide\a*\l*\k*\*ths\r*\a*\advance\k*-\r*%
\dt*=\the\k*\L*}
\def\arcto#1(#2,#3)[#4]{\rlap{\toks0={#2}\toks1={#3}\calcnum*#1(#2,#3)%
\dm*=135sp\dm*=#1\dm*\d**=#1\Lengthunit\ifdim\dm*<0pt\dm*-\dm*\fi%
\multiply\dm*\r*\a*=.3\dm*\a*=#4\a*\ifdim\a*<0pt\a*-\a*\fi%
\advance\dm*\a*\N*\dm*\divide\N*10000%
\divide\N*2\multiply\N*2\advance\N*1%
\L*=-.25\d**\L*=#4\L*\divide\d**\N*\divide\L*\*ths%
\m*\N*\divide\m*2\dm*=\the\m*5sp\l*\dm*%
\sm*\n*\*one\loop\calcparab*\shl**{-\dt*}%
\advance\n*1\ifnum\n*<\N*\repeat}}
\def\arrarcto#1(#2,#3)[#4]{\L*=#1\Lengthunit\L*=.54\L*%
\arcto#1(#2,#3)[#4]\mov*(#2\L*,#3\L*){\d*=.457\L*\d*=#4\d*\d**-\d*%
\mov*(#3\d**,#2\d*){\arrow.02(#2,#3)}}}
\def\dasharcto#1(#2,#3)[#4]{\rlap{\toks0={#2}\toks1={#3}\calcnum*#1(#2,#3)%
\dm*=\the\N*5sp\a*=.3\dm*\a*=#4\a*\ifdim\a*<0pt\a*-\a*\fi%
\advance\dm*\a*\N*\dm*%
\divide\N*20\multiply\N*2\advance\N*1\d**=#1\Lengthunit%
\L*=-.25\d**\L*=#4\L*\divide\d**\N*\divide\L*\*ths%
\m*\N*\divide\m*2\dm*=\the\m*5sp\l*\dm*%
\sm*\n*\*one\loop%
\calcparab*\shl**{-\dt*}\advance\n*1%
\ifnum\n*>\N*\else\calcparab*%
\sh*(#2,#3){\kern#3\dt*\lower#2\dt*\hbox{\sm*}}\fi%
\advance\n*1\ifnum\n*<\N*\repeat}}
\def\*shl*#1{%
\c*=\the\n*\d**\advance\c*#1\a**\d*\dt*\advance\d*#1\b**%
\a*=\the\toks0\c*\b*=\the\toks1\d*\advance\a*-\b*%
\b*=\the\toks1\c*\d*=\the\toks0\d*\advance\b*\d*%
\raise\b*\rlap{\kern\a*\unhcopy\spl*}}
\def\calcnormal*#1{%
\b**=10000sp\a**\b**\k*\n*\advance\k*-\m*%
\multiply\a**\k*\divide\a**\m*\a**=#1\a**\ifdim\a**<0pt\a**-\a**\fi%
\ifdim\a**>\b**\d*=.96\a**\advance\d*.4\b**%
\else\d*=.96\b**\advance\d*.4\a**\fi%
\d*=.01\d*\r*\d*\divide\a**\r*\divide\b**\r*%
\ifnum\k*<0\a**-\a**\fi\d*=#1\d*\ifdim\d*<0pt\b**-\b**\fi%
\k*\a**\a**=\the\k*\dd*\k*\b**\b**=\the\k*\dd*}
\def\wavearcto#1(#2,#3)[#4]{\rlap{\toks0={#2}\toks1={#3}\calcnum*#1(#2,#3)%
\c*=\the\N*5sp\a*=.4\c*\a*=#4\a*\ifdim\a*<0pt\a*-\a*\fi%
\advance\c*\a*\N*\c*\divide\N*20\multiply\N*2\advance\N*-1\multiply\N*4%
\d**=#1\Lengthunit\dd*=.012\d**\ifdim\d**<0pt\d**-\d**\fi\L*=.25\d**%
\divide\d**\N*\divide\dd*\N*\L*=#4\L*\divide\L*\*ths%
\m*\N*\divide\m*2\dm*=\the\m*0sp\l*\dm*%
\sm*\n*\*one\loop\calcnormal*{#4}\calcparab*%
\*shl*{1}\advance\n*\*one\calcparab*%
\*shl*{1.3}\advance\n*\*one\calcparab*%
\*shl*{1}\advance\n*2%
\dd*-\dd*\ifnum\n*<\N*\repeat\n*\N*\shl**{0pt}}}
\def\triangarcto#1(#2,#3)[#4]{\rlap{\toks0={#2}\toks1={#3}\calcnum*#1(#2,#3)%
\c*=\the\N*5sp\a*=.4\c*\a*=#4\a*\ifdim\a*<0pt\a*-\a*\fi%
\advance\c*\a*\N*\c*\divide\N*20\multiply\N*2\advance\N*-1\multiply\N*2%
\d**=#1\Lengthunit\dd*=.012\d**\ifdim\d**<0pt\d**-\d**\fi\L*=.25\d**%
\divide\d**\N*\divide\dd*\N*\L*=#4\L*\divide\L*\*ths%
\m*\N*\divide\m*2\dm*=\the\m*0sp\l*\dm*%
\sm*\n*\*one\loop\calcnormal*{#4}\calcparab*%
\*shl*{1}\advance\n*2%
\dd*-\dd*\ifnum\n*<\N*\repeat\n*\N*\shl**{0pt}}}
\def\hr*#1{\clap{\xL*=\xscale\Lengthunit\vrule width#1\xL* height.1pt}}
\def\shade#1[#2]{\rlap{\Lengthunit=#1\Lengthunit%
\smov(0,#2.05){\hr*{.994}}\smov(0,#2.1){\hr*{.980}}%
\smov(0,#2.15){\hr*{.953}}\smov(0,#2.2){\hr*{.916}}%
\smov(0,#2.25){\hr*{.867}}\smov(0,#2.3){\hr*{.798}}%
\smov(0,#2.35){\hr*{.715}}\smov(0,#2.4){\hr*{.603}}%
\smov(0,#2.45){\hr*{.435}}}}
\def\dshade#1[#2]{\rlap{%
\Lengthunit=#1\Lengthunit\if#2-\def\t*{+}\else\def\t*{-}\fi%
\smov(0,\t*.025){%
\smov(0,#2.05){\hr*{.995}}\smov(0,#2.1){\hr*{.988}}%
\smov(0,#2.15){\hr*{.969}}\smov(0,#2.2){\hr*{.937}}%
\smov(0,#2.25){\hr*{.893}}\smov(0,#2.3){\hr*{.836}}%
\smov(0,#2.35){\hr*{.760}}\smov(0,#2.4){\hr*{.662}}%
\smov(0,#2.45){\hr*{.531}}\smov(0,#2.5){\hr*{.320}}}}}
\def\vdot{\rlap{\kern-1.9pt\lower1.8pt\hbox{$\scriptstyle\bullet$}}}
\def\vtimes{\rlap{\kern-3pt\lower1.8pt\hbox{$\scriptstyle\times$}}}
\def\vDot{\rlap{\kern-2.3pt\lower2.7pt\hbox{$\bullet$}}}
\def\vTimes{\rlap{\kern-3.6pt\lower2.4pt\hbox{$\times$}}}
\catcode`\*=12
\newcount\CatcodeOfAtSign
\CatcodeOfAtSign=\the\catcode`\@
\catcode`\@=11
\newcount\n@ast
\def\n@ast@#1{\n@ast0\relax\get@ast@#1\end}
\def\get@ast@#1{\ifx#1\end\let\next\relax\else%
\ifx#1*\advance\n@ast1\fi\let\next\get@ast@\fi\next}
\newif\if@up \newif\if@dwn
\def\up@down@#1{\@upfalse\@dwnfalse%
\if#1u\@uptrue\fi\if#1U\@uptrue\fi\if#1+\@uptrue\fi%
\if#1d\@dwntrue\fi\if#1D\@dwntrue\fi\if#1-\@dwntrue\fi}
\def\halfcirc#1(#2)[#3]{{\Lengthunit=#2\Lengthunit\up@down@{#3}%
\if@up\smov(0,.5){\arc[-][-]\arc[+][-]}\fi%
\if@dwn\smov(0,-.5){\arc[-][+]\arc[+][+]}\fi%
\def\lft{\smov(0,.5){\arc[-][-]}\smov(0,-.5){\arc[-][+]}}%
\def\rght{\smov(0,.5){\arc[+][-]}\smov(0,-.5){\arc[+][+]}}%
\if#3l\lft\fi\if#3L\lft\fi\if#3r\rght\fi\if#3R\rght\fi%
\n@ast@{#1}%
\ifnum\n@ast>0\if@up\shade[+]\fi\if@dwn\shade[-]\fi\fi%
\ifnum\n@ast>1\if@up\dshade[+]\fi\if@dwn\dshade[-]\fi\fi}}
\def\halfdashcirc(#1)[#2]{{\Lengthunit=#1\Lengthunit\up@down@{#2}%
\if@up\smov(0,.5){\dasharc[-][-]\dasharc[+][-]}\fi%
\if@dwn\smov(0,-.5){\dasharc[-][+]\dasharc[+][+]}\fi%
\def\lft{\smov(0,.5){\dasharc[-][-]}\smov(0,-.5){\dasharc[-][+]}}%
\def\rght{\smov(0,.5){\dasharc[+][-]}\smov(0,-.5){\dasharc[+][+]}}%
\if#2l\lft\fi\if#2L\lft\fi\if#2r\rght\fi\if#2R\rght\fi}}
\def\halfwavecirc(#1)[#2]{{\Lengthunit=#1\Lengthunit\up@down@{#2}%

if@up\smov(0,.5){\wavearc[-][-]\wavearc[+][-]}\fi%
\if@dwn\smov(0,-.5){\wavearc[-][+]\wavearc[+][+]}\fi%
\def\lft{\smov(0,.5){\wavearc[-][-]}\smov(0,-.5){\wavearc[-][+]}}%
\def\rght{\smov(0,.5){\wavearc[+][-]}\smov(0,-.5){\wavearc[+][+]}}%
\if#2l\lft\fi\if#2L\lft\fi\if#2r\rght\fi\if#2R\rght\fi}}
\def\Circle#1(#2){\halfcirc#1(#2)[u]\halfcirc#1(#2)[d]\n@ast@{#1}%
\ifnum\n@ast>0\clap{%
\dimen0=\xscale\Lengthunit\vrule width#2\dimen0 height.1pt}\fi}
\def\wavecirc(#1){\halfwavecirc(#1)[u]\halfwavecirc(#1)[d]}
\def\dashcirc(#1){\halfdashcirc(#1)[u]\halfdashcirc(#1)[d]}
\def\xscale{1}
\def\yscale{1}
\def\Ellipse#1(#2)[#3,#4]{\def\xscale{#3}\def\yscale{#4}%
\Circle#1(#2)\def\xscale{1}\def\yscale{1}}
\def\dashEllipse(#1)[#2,#3]{\def\xscale{#2}\def\yscale{#3}%
\dashcirc(#1)\def\xscale{1}\def\yscale{1}}
\def\waveEllipse(#1)[#2,#3]{\def\xscale{#2}\def\yscale{#3}%
\wavecirc(#1)\def\xscale{1}\def\yscale{1}}
\def\halfEllipse#1(#2)[#3][#4,#5]{\def\xscale{#4}\def\yscale{#5}%
\halfcirc#1(#2)[#3]\def\xscale{1}\def\yscale{1}}
\def\halfdashEllipse(#1)[#2][#3,#4]{\def\xscale{#3}\def\yscale{#4}%
\halfdashcirc(#1)[#2]\def\xscale{1}\def\yscale{1}}

\def\halfwaveEllipse(#1)[#2][#3,#4]{\def\xscale{#3}\def\yscale{#4}%
\halfwavecirc(#1)[#2]\def\xscale{1}\def\yscale{1}}
\catcode`\@=\the\CatcodeOfAtSign

\begin{document}
\def\theequation{\arabic{section}.\arabic{equation}} 
\begin{titlepage} 
\begin{flushright} 
JINR E2-99-190 \\
hep-th/9908054\\ 
August 1999 
\end{flushright} 
 
\vskip1.5cm \centerline{\large\bf HOLOMORPHIC EFFECTIVE ACTION 
OF N=2 SYM THEORY} 

\vskip0.3cm
\centerline{\large\bf FROM HARMONIC SUPERSPACE WITH CENTRAL CHARGES}  
 
\vskip1.5cm 
\centerline{ {\large\bf S. Eremin}, $\quad $ {\large\bf E. Ivanov} } 
 
\vskip.6cm 
\centerline{\it $\;$ Bogoliubov Laboratory of Theoretical Physics,} 
\centerline{\it $\;$ JINR, 141 980 Dubna, Moscow Region,} 
\centerline{\it $\;$ Russian Federation}

\vskip1cm

\begin{abstract} 
\noindent We compute the one-loop holomorphic effective action of the 
massless Cartan sector of $N=2$ SYM theory in the Coulomb 
branch, taking into account the contributions both from 
the charged hypermultiplets and off-diagonal components 
of the gauge superfield. We use the manifestly supersymmetric 
harmonic superfields diagram techniques adapted to $N=2$ supersymmetry 
with the central charges induced by Cartan generators. 
The (anti)holomorphic part proves to be proportional 
to the central charges and it has the generic form of Seiberg's action 
obtained by integrating $U(1)_R$ anomaly. 
It vanishes for $N=4$ SYM theory, i.e. 
the coupled system of $N=2$ gauge superfield and hypermultiplet in 
the adjoint representation.
\end{abstract} 
 
\end{titlepage}  

\section{Introduction} 
Quantum calculations in $N=2$ and $N=4$ super Yang-Mills (SYM) 
theories within the $N=2$ harmonic superspace (HSS) 
approach \cite{gios2,gios1} were a subject of several recent studies 
\cite{ab1}-\cite{nonhol}. The main advantage of the HSS approach is that it 
offers a unique opportunity to keep manifest off-shell 
$N=2$ SUSY at all steps of computation (as distinct, e.g., 
from the $N=1$ superfield techniques). 

It also allows to clearly see that 
the genuine $N=2$ SUSY in the Coulomb branch 
of $N=2$ gauge theory (including matter hypermultiplets) \cite{SW} 
is the central charge-extended $N=2$ SUSY, with the central charges 
induced by non-zero vacuum values of the diagonal (Cartan) components 
of the gauge superfield 
\cite{ikz,bbik}. These vacuum values simultaneously trigger a 
spontaneous breaking of gauge symmetry down to its Cartan torus, 
producing BPS masses for all superfields with non-trivial charges 
with respect to the Cartan generators, including 
off-diagonal components of the gauge superfield.
This modification of $N=2$ SUSY implies that in the course of quantum 
computations in $N=2$ gauge theories in the Coulomb branch 
one should use the HSS techniques (propagators, etc) pertinent 
just to the central charge-extended 
case. They are a proper generalization of the quantum 
HSS methods worked out in \cite{gios2,gios1} for ordinary $N=2$ SUSY. 

In \cite{ikz} these techniques were applied to compute 
the leading contributions 
to the low-energy Wilsonian effective action of hypermultiplets 
minimally coupled to $U(1)$ gauge supermultiplet. The 
hypermultiplets acquire BPS masses on account of 
non-zero vacuum values of the gauge superfield which 
induce central charges in $N=2$ superalgebra. The HSS quantum 
computation with the central charge-massive hypermultiplet propagator 
showed that the leading term 
is uniquely defined as an integral over the analytic harmonic superspace with 
the quartic self-coupling of the hypermultiplet superfields as 
the Lagrangian density. This gives rise to an interesting 
consequence for the physical bosonic fields 
of the hypermultiplet: their metric ceases to be flat and 
is quantum-mechanically deformed into a non-trivial  
hyper-K\"ahler metric. It is the Taub-NUT metric in the case 
of one hypermultiplet, or its higher-dimensional generalizations 
in the case of few hypermultiplets. Simultaneously, a scalar 
fields potential arises. 
The same approach was used in \cite{bbik} to demonstrate 
the central-charge origin of 
the Seiberg-type \cite{seib} perturbative holomorphic contributions 
to the effective action of gauge superfield in the same system 
of $U(1)$ gauge supermultiplet interacting
with charged hypermultiplets. Once again, 
these contributions were found to be proportional 
to the central charges of $N=2$ superalgebra, viz. 
vacuum values of $U(1)$ gauge 
superfields\footnote{This calculation was firstly made within 
the standard quantum HSS approach, by treating non-zero vacuum 
value of the gauge superfield 
as a perturbation \cite{ab1}.}. 
Thus, the same mechanism of non-zero central charges proves 
to be responsible for the two types of analytic contributions to 
 the low-energy effective action in the Coulomb branch: 
the complex-analytic (holomorphic) ones in the gauge fields sector and 
the harmonic-analytic ones in the hypermultiplet sector. It is the 
quantum HSS approach that makes manifest the common origin 
of these two different sorts of the analytic corrections.

In all these studies, only contributions of charged hypermultiplets 
were taken into account, i.e. the gauge theory was assumed to be abelian.
On the other hand, it is natural to start from 
the full non-abelian $N=2$ gauge theory and to take account of 
the contributions of the charged components of the gauge superfield 
itself, along with those of hypermultiplets. 
The appearance of the holomorphic effective action for 
the massless Cartan gauge superfields in the pure $N=2$ SYM theory 
(as well as non-holomorphic corrections to it) in the HSS approach 
was discussed in \cite{nonhol} using the background fields 
method \cite{bfm}. It would be of interest 
to compute these contributions directly, by the same methods 
as in \cite{ikz,bbik}, i.e. by treating the central charge-massive 
off-diagonal components of the gauge superfield on equal footing 
with the massive hypermultiplets. 

This is what we do in the present paper. In Sect.~2, 3 
we briefly discuss 
the quantum techniques of the central charge-extended HSS 
in application to $N=2$ SYM theory. We give the relevant Lagrangians, 
propagators and vertices and demonstrate that at one loop only three-linear 
vertex from the infinite sequence of $N=2$ SYM vertices should 
actually be taken into account while computing the 
holomorphic part of the effective 
action. Then, in Sect.~4, choosing for simplicity 
$SU(2)$ as the gauge group, 
we calculate the one-loop holomorphic correction to the action 
of the neutral component of the gauge superfield. 
It comes from the box-type diagrams with the neutral 
component on the external legs and the massive charged components 
(together with the 
charged components of the Faddeev-Popov ghosts) running inside. 
The answer coincides with the Seiberg's effective action obtained 
in \cite{seib} by integrating $U(1)_R$ anomaly. 
We compare it with the contribution of 
$q$-hypermultiplet in the adjoint representation and find both 
to cancel each other, in agreement 
with the absence of holomorphic terms in the effective action of 
$N=4$ SYM to which this $N=2$ system amounts 
(actually, this is a modified $N=4$ SYM theory, with 
a central-charges extended $N=4$ SUSY). The calculations can be easily 
generalized to an arbitrary semi-simple gauge group. The corresponding 
holomorphic contribution is given by a sum over positive roots 
in the spirit of refs. \cite{roots}. 
In two Appendices some technical points are treated. 
In particular, we check that our results are independent 
of the choice of $\alpha$ -gauges in the massive gauge 
superfield propagators. 

\setcounter{equation}{0}

\section{Classical and microscopic N=2 SYM actions 
in the presence of central charges} 

The classical HSS action of $N=2$ SYM theory minimally coupled 
to hypermultiplets reads 
\bea
\label{s}
S=S^{N=2}_{SYM} + S_{HP}^{N=2}\;.
\eea
Here \cite{BMZ1}
\bea
\label{sym}
S^{N=2}_{SYM}=\frac{1}{2g^2} \mbox {Tr} \int{\rm d}^{12}z\sum_{n=2}^{\infty}\frac{(-i)^n}
{n}\int{\rm d}u_1\dots{\rm d}u_n\,\frac{V^{++}(z,u_1)\dots V^{++}(z,u_n)}
{(u^+_1u^+_2)(u^+_2u^+_3)\dots (u^+_nu^+_1)}
\eea
and $V^{++}(z,u) \equiv V^{++}_A(z,u)T_{A}$ is the $N=2$ analytic 
gauge potential \footnote{The generators $T_A$ of the gauge group $G$ 
are chosen Hermitean, 
$[T_A,T_B]=if_{ABC}T_C$, and belonging to the adjoint representation. 
We normalize them so that ${\rm Tr}\;(T_AT_B)=\delta_{AB}\;$.}. 
The integration in (\ref{sym}) goes over 
the full $N=2, \,D=4$ HSS in the central basis, 
the harmonic distributions and 
integrals are defined in \cite{gios1}. 
The second term in (\ref{s}) is the hypermultiplet action
\bea
S_{HP}^{N=2}=\frac{1}{2}\int 
d\zeta^{(-4)} q^{+B'}_a\nabla^{++}q^{+B'a}\quad,
\nabla^{++}q^{+B'a} = D^{++}q^{+B'a} + ig V^{++ D}(t_D)^{B'C'}q^{+C'a}\;.
\eea
Here the integration goes over the harmonic analytic superspace 
$\{\zeta\} \equiv \{z_A^m, \theta^+_\alpha, \bar\theta^+_{\dot \alpha}, 
u^{\pm}_i\}$, 
$q^{+ B'a}(\zeta)$ are the analytic hypermultiplet superfields 
in a representation $R$ of the group $G$, $(t_D)^{B'C'}$ are 
the corresponding generators. We shall mainly deal with 
$q^+$ in the adjoint representation, i.e. $q^{+Ba}$. In this case 
the action (\ref{s}) is that of $N=4$ SYM theory 
in the $N=2$ superfield formulation, with a hidden second 
$N=2$ SUSY \cite{gios1}. 

We shall concentrate on the first term in~(\ref{s}) 
which is the pure $N=2$ SYM action.
Let $C$ be the Cartan subgroup of the gauge group $G$. 
We shall use the decomposition
$$V^{++}=V^{++}_{(c)}+ V^{++}_{G/C}, 
$$ 
where $V^{++}_{(c)}$ takes values in the Lee algebra 
of $C$ and $V^{++}_{G/C}$ belongs 
to the coset $G/C$. In the standard Cartan-Weyl basis $V^{++}_{(c)}$ 
and $V^{++}_{G/C}$ collect, respectively, the diagonal 
and off-diagonal components of $V^{++}$. 
In what follows we assume that all Cartan components 
develop non-zero vacuum expectation values 
$v^{++} \equiv <V^{++}_{(c)}>$:
\bea
\label{redef}
V^{++}=v^{++}+{\widetilde{V}}^{++}=-(\theta^+\theta^+)\bar{W}_0
-(\bar\theta^+\bar\theta^+)W_0+{\widetilde{V}}^{++}\quad ,
\qquad<{\widetilde{V}}^{++}>=0\;.
\eea
Here ${\bar{W}}_0, W_0$ are some constants (moduli) with values 
in the Lee algebra of $C$. This corresponds to the Coulomb 
branch of the theory, with the original 
gauge symmetry being spontaneously broken down to that 
with respect to the Cartan subgroup. The leading terms of 
the Cartan subalgebra valued superfield strengths 
$W_\cc \equiv W_{(c)} - W_0$~, $\bar W_\cc$  
are expressed through the potential 
$V^{++}_\cc \equiv \tilde{V}^{++}_{(c)}$ as
\footnote{
We use the following notation: $(D)^2 = 
\frac{1}{4}{D}^\alpha{D}_\alpha = \frac{1}{4}(D D)$, $(\bar D)^2 = 
{1\over 4}\bar D_{\dot \alpha} \bar D^{\dot \alpha} = 
{1\over 4}(\bar D \bar D)$, 
$(D)^4 = (D)^2(\bar D)^2$.} 
\bea
\label{W_Vc}
W_\cc=-\int{\rm d}u({\bar{D}}^-)^2V_\cc^{++}\;,\qquad
{\bar{W}}_\cc=-\int{\rm d}u(D^-)^2V_\cc^{++}\;.
\eea
Full such strengths include, of course, nonlinear terms 
$\sim {\widetilde{V}}^{++}_{G/C} \equiv \hat V^{++}$. 
In our computation of the Cartan subgroup sector of 
the effective action of $N=2$ SYM we shall be interested 
only in such leading terms, assuming that 
the full strengths are restored by non-abelian 
gauge invariance.    

The non-vanishing background superfield $v^{++}$ triggers 
the spontaneous breakdown of $G$ to $C$ and simultaneously 
generates constant $U(1)$ central charges 
in $N=2$ superalgebra \cite{ikz,BMZ2}. 
They explicitly break $U(1)_R$ automorphism 
symmetry of the original $N=2$ SUSY.
The generators of the central charges-modified $N=2$ superalgebra are 
\bea
{\cal Q}^k_\alpha=Q^k_\alpha+i\theta^k_\alpha\bar{W}_0\quad , \quad
{\bar{\cal Q}}_{k\dot\alpha}=
\bar{Q}_{k\dot{\alpha}}+
i\bar{\theta}_{k\dot{\alpha}}{W_0}\;,
\eea
and the algebra of the flat covariant derivatives is modified as 
\bea
\label{derev}
&&\{{\cal D}^+_\alpha,{\cal D}^-_\beta\}=
-2i\epsilon_{\alpha\beta}\bar{W}_0\qquad
\{\bar{{\cal D}}^+_{\dot\alpha},\bar{{\cal D}}^-_{\dot\beta}\}=
2i\epsilon_{\dot{\alpha}\dot{\beta}}W_0 \nn
&&\{\bar{{\cal D}}^+_{\dot\alpha},{\cal D}^-_\beta\}=
-\{ {\cal D}^+_\beta ,\bar{{\cal D}}^-_{\dot\alpha}   \}
=2i{\cal D}_{\dot\alpha\beta} \nn
&&[{\cal D}^{++},
 {\cal D}_\alpha^-]
={\cal D}^+_{\alpha}\qquad
[{\cal D}^{++},\bar{\cal D}^-_{\dot\alpha}]=\bar{\cal D}^+_{\dot\alpha} \nn
&&\{{\cal D}^+_\alpha,{\cal D}^+_\beta\}=
\{\bar{\cal D}^+_{\dot{\alpha}},\bar{\cal D}^+_{\dot{\beta}}\}=0\;.
\eea
This algebra can be interpreted as the algebra of the
full covariant $N=2$ SYM derivatives in a covariantly constant gauge
background $v^{++}$ \cite{bfm,ikz}. Then, following \cite{gios2}, 
we can introduce the ``bridge'' $v$
corresponding to the particular analytic gauge potential $v^{++}$
\bea
v=-(\theta^+\theta^-)\bar{W}_0
-(\bar\theta^+\bar\theta^-)W_0\;, \quad v^{++} = D^{++}v\;.
\eea
It allows one to choose different frames for the harmonic superfields 
of the central-charge extended $N=2$ supersymmetry, with 
the manifest or ``covariant'' 
harmonic analyticities, making a finite Cartan subgroup transformation 
with $v$ as the group parameter. In one of these frames, 
the ``$\lambda $-frame'', 
the derivatives ${\cal D}^+_\alpha$, $\bar{\cal D}^+_{\dot\alpha}$ 
contain no 
central charge terms, that is $({\cal D}^+_\alpha)_{\lambda} = D^+_\alpha$, 
$(\bar{\cal D}^+_{\dot\alpha})_\lambda = \bar D^+_{\dot\alpha}$. 
In this frame, 
like in the case of ordinary $N=2$ SUSY,  
one can define the manifestly analytic harmonic superfields 
which depend only on 
half of the original Grassmann coordinates. At the same time, 
the harmonic derivatives 
$(D^{\pm\pm})_\lambda$ necessarily contain the central-charge terms. 
On the 
contrary, in the ``$\tau$-frame'' the above spinor derivative include the 
central-charge terms, while the harmonic derivatives do not. 
In this frame 
the harmonic analyticity is covariant. The relation between the two frames 
is given by   
\bea
(\widetilde{V}^{++})_\tau=
e^{-iv}(\widetilde{V}^{++})_{\lambda}e^{iv} ,\qquad 
({\cal D})_\tau=
e^{-iv}({\cal D})_{\lambda}e^{iv}, \label{transf} 
\eea  
where ${\cal D}$ stands for the spinor or harmonic derivative. 
For what follows, it will be 
useful to give the $\tau$ frame derivatives 
in the central basis of $N=2$ HSS
(for brevity, we omit the subscript $\tau$)
\bea
{\cal D}^{+}_\alpha=D^+_\alpha-i\theta^+_\alpha{\bar{W}}_0\,,\quad
{\bar{{\cal D}}}^{+}_{\dot{\alpha}}=\bar{D}^+_{\dot\alpha}
-i{\bar\theta}^+_{\dot\alpha}{W}_0\,,\quad
{\cal D}^{\pm\pm}=u^{\pm i}\frac{\partial}{\partial u^{\mp i}}\;.
\eea   
The algebra (\ref{derev}) does not depend 
on the choice of frame and/or basis.

The realization of $N=2$ SUSY preserves 
its old form only on the Cartan part of the shifted
superfield ${\widetilde{V}}^{++}$, since the latter is evidently 
invariant under rigid $C$ transformations. It is modified 
on all other superfields having non-trivial charges 
with respect to $C$ and, hence, non-zero central charges. 
All these superfields acquire BPS masses by the standard 
Scherk-Schwarz mechanism. This concerns as well the off-diagonal 
component $\widetilde{V}^{++}_{G/C} \equiv \hat V^{++}$ of 
$\tilde{V}^{++}$. 
One of the two real $G$-algebra-valued physical scalar 
fields in it, along with the physical fermions, become massive 
by the same mechanism. 
The remaining physical dimension scalar field 
is the $G/C$ Goldstone boson and it produces 
a mass for the $G/C$ gauge fields via the Higgs effect 
(this scalar field fully disappears in the ``unitary'' gauge). 
As a result, only the physical fields contained in 
$\tilde{V}^{++}_{(c)}$ remain massless (for the gauge group $G$ of rank 
$r$ these are $r$ complex scalar fields, $r$ gauge fields and 
$2r$ gaugini Weyl spinors). 
   
Let us return to the $N=2$ SYM action~(\ref{sym}). 

The central-charge 
induced splitting~(\ref{redef}) of $\tilde{V}^{++}$ 
into the massless and massive pieces 
$\widetilde{V}^{++}_\ccc \equiv V^{++}_\cc$ and $\hat{V}^{++}$, 
due to the shift $v^{++}$, essentially redefines the kinetic 
part of the action. In particular, a mass term for the charged components
$\hat{V}^{++}$ is generated  
(this phenomenon is analogous to 
the central-charge modification of the free $q^+$ action, 
see ref. \cite{ikz,bbik}). 
The precise form of the split action can be found using the fact that 
the redefinition ~(\ref{redef}) and the relations ~(\ref{transf}) 
are a particular case of the splitting of $V^{++}$ into the quantum 
and background parts within 
the $N=2$ background fields method \cite{bfm}, with $v^{++}$ and 
$\widetilde{V}^{++}$ playing the roles of the background and quantum parts.   
Then,  following \cite{bfm}, we rewrite (\ref{sym}) as 
\bea
\label{tausym}
S_{SYM}(\widetilde{V}^{++}+v^{++}) &=&
S(v^{++})+ \frac{(-i)}{2}\,\mbox{Tr}
\int{\rm d}^{12}z{\rm d}u\,({\widetilde{V}}^{++})_\tau
e^{iv}{\it D}^{--}e^{-iv} \nn
&+& \frac{1}{2g^2}\mbox{Tr}\int{\rm d}^{12}z
\sum_{n=2}^{\infty}\frac{(-i)^n}
{n}\int{\rm d}u_1\dots{\rm d}u_n\,\frac{({\widetilde{V}}^{++}_1)_\tau\dots
({\widetilde{V}}^{++}_n)_\tau}
{(u_1^+u_2^+)\dots (u_n^+u_1^+)}.
\eea
First two terms in~(\ref{tausym}) are of no relevance for 
the further consideration and can be omitted (actually, the second 
term is zero as a consequence of the manifest analyticity 
of $\widetilde{V}^{++}$ and the fact that $v^{++}, v$ are 
defined on the commutative Cartan subalgebra).
This resulting form of $N=2$ SYM action is the convenient 
starting point for constructing full quantum version of the 
central-charge modified theory.

The Faddeev-Popov-t'Hooft procedure changes insignificantly
for the massive case as compared to the case of pure massless $V^{++}$ 
worked out in \cite{gios1}. 
The main difference is related to the systematic use of the algebra 
of the centrally-extended covariant derivatives ~(\ref{derev}). 
Following the same steps as in \cite{gios1}, 
the gauge-fixed quadratic part of 
the split $N=2$ SYM action, together with the Faddeev-Popov ghosts part, can 
be written as (for the standard $\alpha$-gauges)   
\bea
S^{(2)}_{SYM}+S_{FP} &=&
S^{(2)}(\widetilde{V}^{++}_\ccc) + \frac{1}{4\alpha}\,\mbox{Tr}
\int{\rm d}\zeta^{(-4)}\,
({\hat{V}}^{++})_\tau\bbx({\hat{V}}^{++})_\tau \nn
&&+\;\frac{1}{4}\,(1+\frac{1}{\alpha})\,\mbox{Tr}
\int{\rm d}\zeta^{(-4)}_1{\rm d}u_2\,
({\hat{V}}^{++}_1)_\tau({\cal D}^+_1)_\tau^4\frac{1}
{(u_1^+u_2^+)^2}({\hat{V}}^{++}_2)_\tau \nn
&&+\;i\,\mbox{Tr}\int{\rm d}\zeta^{(-4)}\,F_\tau{\it D}^{++}\{{\it D}^{++}P_\tau+
[({\widetilde{V}}^{++})_\tau\,,\, P_\tau]\}\;.\label{micro}
\eea
Here $\bbx=\Box + W_0{\bar{W}}_0$ and $S^{(2)}(\widetilde{V}^{++}_\ccc)$ is 
the standard quadratic action for the massless $C$ algebra 
valued superfields. The latter has the same form 
as that of $\hat{V}^{++}$, but with taking the purely flat expressions 
for the Box operator and spinor derivatives (with 
no terms $\sim W_0, \bar{W}_0$ producing mass for $\hat{V}^{++}$). 
Notice the appearance of the 
index $\tau$ in the FP action: the ghost superfields 
$F$ and $P$ are in the adjoint representation like $\widetilde{V}^{++}$ 
and therefore their off-diagonal components acquire non-trivial 
central charges and become massive like the charged $q^+$ hypermultiplets. 
We also note that, though $(\hat V^{++})_\tau, F_\tau, P_\tau $ 
are {\it covariantly} analytic, the integrands in (\ref{micro}) are 
{\it manifestly} analytic: this is because 
they are invariant under the global 
$G$ symmetry and, hence, are the central charge singlets. 
    
The full microscopic quantum action for the central-charges extended 
system of $N=2$ SYM and hypermultiplet superfields 
can be obtained as a sum of ~(\ref{micro}), the self-interaction 
$N=2$ SYM part and the hypermultiplet action    
\be
S_{micr} = S^{(2)}_{SYM} + S^{int}_{SYM} + S_{FP} + S^{N=2}_{HP}~, 
\label{micro2}
\ee
where 
\bea
S^{int}_{SYM}=
&&\frac{1}{2g^2}\mbox{Tr}\int{\rm d}^{12}z\sum_{n=3}^{\infty}\frac{(-i)^n}
{n}\int{\rm d}u_1\dots{\rm d}u_n\,\frac{({\widetilde{V}}^{++}_1)_\tau\dots
({\widetilde{V}}^{++}_n)_\tau}
{(u_1^+u_2^+)(u_2^+u_3^+)\dots (u_n^+u_1^+)}\;. \label{selfN2}
\eea
and \cite{ikz}
\bea
&& S_{HP}^{N=2}=\frac{1}{2}\int d\zeta^{(-4)} (q^{+B'}_a)_\tau\tilde{\nabla}^{++}
(q^{+B'a})_\tau \quad, \label{hyp2}\nonumber \\
&& \tilde{\nabla}^{++}(q^{+B'a})_\tau = D^{++}(q^{+B'a})_\tau + 
ig (\tilde{V}^{++ D})_\tau(t_D)^{B'C'}(q^{+C'a})_\tau\;. 
\eea
It is easy to show that action (\ref{micro2}) 
is invariant under the appropriate 
BRST transformations which replace the gauge ones in the quantum case. 
These are a straightforward extension of those for the massless case 
\cite{gios1}. 

\setcounter{equation}{0}
\section{Feynman diagram techniques in HSS with central charges}

Now it is easy to read off the Feynman rules. 
It is more convenient to construct them using covariantly
analytic superfields and then to pass to the manifestly analytic ones. The
covariant analytic Feynman rules do not contain any new types of the
vertices.

The Green function for the massless 
Cartan component $\tilde{V}^{++}_{(c)}$ 
has the standard form (see \cite{gios1}). 
The equation defining the Green function 
for the off-diagonal massive $G/C$ components of 
the full  ${\widetilde{V}}^{++}$ directly 
follows from (\ref{micro})
\bea
\label{gren}
\frac{1}{2\alpha}\bbx{\hat{G}}^{(2,2)}(1\mid{3})+
\frac{1}{2}(1+\frac{1}{\alpha})
({\cal D}^+_1)^4\int{\rm d}u_2\frac{1}{(u_1^+u_2^+)^2}
{\hat{G}}^{(2,2)}(z,u_2\mid{3})=\delta_{(A)}^{(2,2)}(1\mid{3})\;,
\eea
where the covariantly analytic $\delta$-function is defined 
by \cite{BMZ2,ikz} 
\bea
\label{dfunction}
[\delta^{(2,2)}_{(A)}(1\mid{2})]_{BC}=
({\cal D}^+_1)^4_{BC}\delta^{12}(z_1-z_2)\delta^{(2,2)}(u_1,u_2)\;.
\eea
The Green function and $\delta$-function are matrices in the adjoint 
representation of the gauge group.

The solution of ~(\ref{gren}) in the Fermi-Feynman gauge $(\alpha=-1)$ is

\vspace{0.5cm}
\ \mov(1,0)\vdot\mov(1,.019){\lin(1.3,0)}\mov(1,-.03){\lin(1.3,0)}
\ \mov(2.13,0)\vdot
\ \mov(2.5,0){$<({\hat{V}}^{++}_1)_\tau\mid({\hat{V}}^{++}_2)_\tau>=
\dis\frac{2}{\bbx_1}({\cal D}^+_1)^4
\delta^{12}(X_1-X_2)\delta^{(-2,2)}(u_1,u_2)\;.
\quad\;(3.3)$}
\vspace{0.5cm}

The ghost propagator is also a direct generalization of the standard one. 
It can be immediately found from the ghost action in (\ref{micro}): 

\vspace{0.5cm}
\ \mov(1,0)\vdot\mov(1,0){\dashlin(1.3,0)}\mov(2.3,0)\vdot
\ \mov(2.5,0){$<(\hat{F}(1))_\tau\mid(\hat{P}(2))_\tau>=
\dis\frac{1}{\bbx_1}({\cal D}^+_1)^4({\cal D}^+_2)^4\delta^{12}(X_1-X_2)
\dis\frac{\dis (u_1^-u_2^-)}{\dis (u_1^+u_2^+)^3}\;.\quad\;(3.4)$}
\vspace{0.5cm}

We also give the expression for the $q^+$ propagator in 
the central-charge case \cite{ikz,bbik}

\vspace{0.5cm}
\ \mov(1,0)\vdot\mov(1,0){\lin(1.3,0)}\mov(2.3,0)\vdot
\ \mov(2.5,0){$<(\hat{q}^+(1))_\tau \mid(\hat{\bar{q}}^+(2))_\tau>=
\dis\frac{1}{\bbx_1}({\cal D}^+_1)^4({\cal D}^+_2)^4\delta^{12}(X_1-X_2)
\dis\frac{\dis{1}}{\dis (u_1^+u_2^+)^3}\;.\;\,(3.5)$}
\vspace{0.3cm}

While constructing the effective action, it is of no actual 
need to know the explicit expressions for vertices in the Feynman rules. 
Nevertheless, for completeness and for further possible use 
we present some of them. 

The vertex $(F)_\tau({\widetilde{V}}^{++})_\tau (P)_\tau$ is

\vspace{0.2cm}
\ \mov(1,0){\dashlin(2,0)}\mov(2,0)\vdot
\ \mov(1.9,0){\trianglin(0,1)}
\ \mov(0.77,0.1){$B$}
\ \mov(1.7,0.9){$A$}
\ \mov(2.3,0.1){$C$}
\ \mov(3.5,0.5){$gf^{ABC}D^{++}_{(B)}\;.\hspace{6.3cm}(3.6)$}
\vspace{0.3cm}

There is an infinite number of the $N=2$ SYM self-interaction vertices, 
as follows from (\ref{selfN2}). We shall present the first two ones 
(actually, it is the three-linear vertex which is of relevance 
for further calculations).

\vspace{0.2cm}
\ \mov(2,0)\vdot
\ \mov(1.9,0){\trianglin(0.8,-0.8)}
\ \mov(1.76,0){\trianglin(-0.8,-0.8)}
\ \mov(1.6,0){\trianglin(0,1)}
\ \mov(3.5,0){$\displaystyle\frac{1}{2}
\frac{igf^{ABC}}{(u_1^+u_2^+)(u_2^+u_3^+)(u_3^+u_1^+)}\;.
\hspace{4.4cm}(3.7)$}
\ \mov(1.4,0.8){$C, u_3$}
\ \mov(.1,-1.){$A, u_1$}
\ \mov(1.4,-1){$B, u_2$}

\vspace{0.5cm}

\ \mov(2,0)\vdot
\ \mov(1.9,0){\trianglin(0.8,.8)}
\ \mov(1.75,0){\trianglin(-0.8,0.8)}
\ \mov(1.5,0){\trianglin(0.8,-.8)}
\ \mov(1.35,0){\trianglin(-0.8,-0.8)}
\ \mov(0.2,1){$D, u_4$}
\ \mov(1.74,1){$C, u_3$}
\ \mov(-0.2,-1.06){$A, u_1$}
\ \mov(1.25,-1.1){$B, u_2$}
\ \mov(3.5,0){$\displaystyle\sum_{sym}\frac{1}{8}\frac{ig^2f^{ABK}f^{KCD}}
{(u_1^+u_2^+)(u_2^+u_3^+)(u_3^+u_4^+)(u_4^+u_1^+)}\;.\hspace{1.2cm}(3.8)$}
\vspace{0.5cm}
\setcounter{equation}{8}

Here the zigzag lines denote the full ${\widetilde{V}}^{++}$ fields.
Below we will deal only with the Cartan component of ${\widetilde{V}}^{++}$
on the external legs. So, to avoid a confusion, we shall denote the latter 
by the wavy lines

\vspace{0.3cm}
\ \mov(1,0)\vdot\mov(1,0){\trianglin(1.3,0)}
\ \mov(2.25,0.13){$\widetilde{V}^{++}$}
\ \mov(5,0)\vdot\mov(5,0){\wavelin(1.3,0)}
 \ \mov(6.25,0.13){$V_\cc^{++}$}

\vspace{0.3cm}

In what follows we shall be interested in the leading contributions to that 
sector of the one-loop $N=2$ SYM effective action which contains only the 
massless SYM superfields ${V}^{++}_\cc$. As we shall show,  
such contributions can come only from the diagrams containing  
the three-linear vertices (3.7) (see Fig. 2). 
Clearly, when the external legs 
are massless ${V}^{++}_\cc$, only 
the massive charged $\hat{V}^{++}$ superfields with 
the propagators (\ref{gren}) can run inside. Thus the sector of $N=2$ SYM 
effective action we are interested in is defined by the 
expression analogous to that 
defining the contribution from the charged 
$q^+$ hypermultiplets \cite{bbik}, 
namely, by  
\bea
\Gamma{}[V^{++}_\cc]=i\,\mbox{Tr}\ln\{\delta^{(4,2)}_A + {V}^{++}_\cc 
\hat{G}^{(2,2)}\}~.
\eea
Here $\delta^{(4,2)}(1|2)_A$ is the appropriate analytic delta-function 
and $\hat{G}^{(2,2)}(1|2)$ is given by (\ref{dfunction}). Notice that 
$({V}^{++}_\cc)_\tau = 
({V}^{++}_\cc)_\lambda = {V}^{++}_\cc$ in view of 
commutativity of $V^{++}_\ccc$ and the bridge $v$ 
(they both belong to the Cartan subalgebra).

Let us argue that the leading holomorphic part of the $N=2$ SYM 
effective action 
in the considered sector can indeed come only from the diagrams 
on Fig. 2, with 
the three-linear vertices. 

The generic argument is as follows.

For a fixed number of 
the external $V^{++}_\cc$ legs any other one-loop diagram 
apart from those on Fig. 2, i.e. a diagram including some number 
of higher-order $N=2$ SYM vertices (of the type depicted on Fig. 1), 
contains not enough spinor derivatives to produce the (anti)holomorphic 
contributions in the local limit (after restoring the full Grassmann 
integration measures at the vertices). 

This important property can be proved in the general case, 
but it is instructive to firstly illustrate it on a simple example. 
The simplest diagram of the above type includes two $\hat{V}^{++}$ 
propagators, one three-linear and one quartic vertices, but its  
contribution is vanishing by the purely algebraic reason related to the 
properties of the gauge group structure constants. 
So we explain the above argument on the example of the supergraph 
depicted on Fig.~1. It could give a non-vanishing contribution to 
the quartic term of the effective action. 

\vspace{0.4cm}
\ \mov(5.3,0){\Ellipse(1.22)[2,1.04]}\mov(5.3,0){\Ellipse(1.18)[2,1]}
\ \mov(6.33,0)\vdot
\ \mov(6.28,0){\wavelin(1,1)}
\ \mov(6.12,0){\wavelin(1,-1)}
\ \mov(3.51,0)\vdot
\ \mov(3.45,0){\wavelin(-1.4,0)}
\ \mov(3.8,0.79){$({\cal D}^+_1)^2{(\bar{\cal D}}^+_1)^2$}
\ \mov(3.7,-0.89){$({\cal D}^+_2)^2{(\bar{\cal D}}^+_2)^2$}
\ \mov(1.5,0.2){$V^{++}_\cc(u_7,{\theta}_2)$}
\ \mov(6.25,0.77){$V^{++}_\cc(u_2,{\theta}_1)$}
\ \mov(6.12,-0.9){$V^{++}_\cc(u_4,{\theta}_1)$}
\ \mov(3.35,-1.8){${\rm Figure\; 1.}$}
\ \mov(4.3,-0.1){$\theta_1$}
\ \mov(2.2,-0.1){$\theta_2$}
\ \mov(4.35,0){\wavelin(1,0)}
\ \mov(5.45,0){$V^{++}_\cc(u_3,{\theta}_1)$}
\vspace{0.4cm}

Using the identities 
\bea
\delta^8(\theta_1-\theta_2){(D)}^n\delta^8(\theta_1-\theta_2)=0
\qquad \mbox{if} \quad n<8 \;, \nn
\qquad \delta^4(\theta_1-\theta_2)({\cal D}_1^+{\cal D}_2^+)
({\cal D}_n^+{\cal D}_n^+)\delta(\theta_1-\theta_2)= 16
(u_1^+u_n^+)(u_2^+u_n^+)\delta(\theta_1-\theta_2) \label{deltaderev}
\eea
this contribution can be put in the form
\bea
&&\int
\frac{{\rm d}^4p_1{\rm d}^4p_2{\rm d}^8{\theta}_1
{\rm d}u_1{\rm d}u_2{\rm d}u_3{\rm d}u_4{\rm d}u_5{\rm d}u_7}
{(2\pi)^8(p_1^2-m^2)(p_2^2-m^2)} \;f(u_1,u_2,u_3,u_4,u_5,u_7) \nn
&&{\times}V^{++}_\cc(\theta_1,u_2)V^{++}_\cc(\theta_1,u_3)V^{++}_\cc(\theta_1,u_4)
V^{++}_\cc(\theta_1,u_7)~,
\eea
with $f(u_1,u_2,u_3,u_4,u_5,u_7)$ being a function 
of the harmonic variables only.
The holomorphic (anti-holomorphic) contribution is the integral over 
$d^4\theta$ ($d^4\bar{\theta}$) with a function of the 
strength $W_\cc$ ($\bar W_\cc$) 
as the integrand. 
Then, taking off the proper number of the derivatives from 
the Grassmann measure, we arrive at the expression 
\bea
&&\int
\frac{{\rm d}^4p_1{\rm d}^4p_2
{\rm d}u_1{\rm d}u_2{\rm d}u_3{\rm d}u_4{\rm d}u_5{\rm d}u_7}
{(2\pi)^8(p_1^2-m^2)(p_2^2-m^2)}
\,f(u_1,u_2,u_3,u_4,u_5,u_7) \nn
&&{\times}\int{\rm d}^4{\theta}
\Big\{({\bar{\cal D}}^+_1)^2({\bar{\cal D}}^-_2)^2
[V^{++}_\cc(\theta_1,u_2)V^{++}_\cc(\theta_1,u_3)V^{++}_\cc(\theta_1,u_4)]
V^{++}_\cc(\theta_1,u_7)]\Big\}\;. 
\label{expr4}
\eea
Recalling the definition of the abelian SF strengths (\ref{W_Vc}), 
we see that one needs at least {\it eight} spinor derivatives to form these 
objects from the four gauge potentials in (\ref{expr4}), while only 
{\it four} such derivatives are present in (\ref{expr4}). 
Thus we conclude that 
no any local holomorphic (anti-holomorphic) contribution can be 
produced by the above supergraph.

In the generic case, every additional external line at some vertex of the 
given supergraph demands two extra spinor derivatives for this diagram  
be capable to produce a holomorphic term in the local limit. 
Such derivatives can be taken off only from the propagators. 
So, for obtaining a holomorphic 
contribution there should be a strict relation
between the numbers of external lines and propagators. 
A simple analysis shows that these numbers should be equal, which 
selects the diagrams on Fig. 2 as the only source of holomorphic contribution 
from the pure $N=2$ SYM to the Cartan sector 
of the effective one-loop $N=2$ SYM action. 

\setcounter{equation}{0}
\section{One-loop holomorphic contribution to the SYM effective \break 
action}

For simplicity, we firstly choose $SU(2)$ as the gauge group, with 
$V^{++}_\cc$ being simply $\tilde{V}^{++}_3T_3$. 
The techniques suitable for 
calculating $n$-particle holomorphic corrections to the effective action 
in the case of $N=2$ SUSY with constant background central charges 
were worked out in \cite{bbik}. Although the consideration in \cite{bbik} 
was limited to the abelian $U(1)$ theory and to the supergraphs with 
$q^+$ hypermultiplets inside, it can be easily extended to our case. 
This is mainly  due to the property that for extracting 
the holomorphic contributions it is sufficient 
to consider the diagrams depicted in Fig. 2. 
The only complication compared to the hypermultiplet case 
is the presence of harmonic non-localities in the pure $N=2$ SYM vertices.

So, in the $N=2$ SYM sector (with no hypermultiplets) 
the holomorphic contribution can come from the diagrams on Fig. 2 
and those with the Faddeev-Popov ghosts inside (Fig. 3). 

\vspace{0.3cm}
\ \mov(2.2,0){\Circle(1.33)}
\ \mov(2,0.67)\vdot
\ \mov(1.9,0.67){\wavelin(0,1)}
\ \mov(1.14,-0.1)\vdot
\ \mov(1.08,-0.1){\wavelin(-1,-0.1)}
\ \mov(2.16,-0.1)\vdot
\ \mov(2.1,-0.1){\wavelin(1,-0.1)}
\ \mov(1.8,-1.)\vdot
\ \mov(1.2,-1.1)\vdot
\ \mov(0.6,-1)\vdot
\ \mov(1.05,1.6){$u_1$}
\ \mov(0.33,0.7){$\om_1$}
\ \mov(0.85,0.7){$\vv_1$}
\ \mov(2.05,-0.02){$u_2$}
\ \mov(1.0,0.15){$\om_2$}
\ \mov(0.78,-0.49){$\vv_2$}
\ \mov(-1.85,-0.04){$u_n$}
\ \mov(-1.1,-0.5){$\om_n$}
\ \mov(-1.31,0.16){$\vv_n$}
\ \mov(-1.15,-1.8){${\rm Figure\;2.}$}
\ \mov(-.704,0){\Circle(1.26)}
\ \mov(4.4,0){\dashcirc(1.3)}
\ \mov(4.2,0.66)\vdot
\ \mov(4.1,0.66){\wavelin(0,1.1)}
\ \mov(3.37,0.22)\vdot
\ \mov(4.49,0.22)\vdot
\ \mov(3.22,0.22){\wavelin(-1.23,0.36)}
\ \mov(4.26,0.22){\wavelin(1.23,0.36)}
\ \mov(3.85,-0.53)\vdot
\ \mov(3.78,-0.53){\wavelin(1.0,-0.7)}
\ \mov(3.1,-0.97)\vdot
\ \mov(2.41,-0.79)\vdot
\ \mov(2.05,-0.3)\vdot
\ \mov(2.87,1.9){$V_\cc^{++}(1)$}
\ \mov(4.64,0.7){$V_\cc^{++}(2)$}
\ \mov(4.15,-1.1){$V_\cc^{++}(3)$}
\ \mov(0.17,0.8){$V_\cc^{++}(n)$}
\ \mov(1.85,-1.8){${\rm Figure\; 3.}$}
\ \mov(-5.55,-0.5){$V_\cc^{++}(n)$}
\ \mov(-1.7,-0.5){$V_\cc^{++}(2)$}
\ \mov(-4.44,1.77){$V_\cc^{++}(1)$}
\vspace{0.3cm}

We begin by computing the first type of contribution. 
It can be represented in the form  
\bea
\label{ym}
\Gamma_{YM}^n[V^{++}]=
&&\frac{(-i)^{n+1}}{n} {\rm Tr}\prod^n_{t=1}
\int
{\rm d}^4x_t{\rm d}^8\theta_t{\rm d}u_t{\rm d}\omega_t{\rm d}\vv_t
\,\frac{V_\cc^{++}(t)}
{(u_t^+\om_t^{+})(\om_t^{+}\vv_t^{+})(\vv_t^{+}u_t)} \nn
&&\times\frac{1}{\Box_t+m^2}
[{\cal D}^+(\theta_t,\vv_t)]^4\delta^{8}(x_t-x_{t+1})
\delta^8(\theta_t-\theta_{t+1})\delta^{(-2,2)}(\vv_t,\om_{t+1})\;. 
\label{start}
\eea
Hereafter, $V_\cc^{++}(t) \equiv V_\cc^{++}(x_t,\theta_t,u_t)$, 
$x_{n+1}\equiv x_1\,,\; u_{n+1}\equiv u_1\,,\; \theta_{n+1}\equiv\theta_1\,,\;
u_{1-1}\equiv u_n\,.$

Passing to the momentum space, we find that
the terms resulting in the holomorphic and anti-holomorphic contribution 
in the low-energy limit are concentrated in the expression
\bea
\Gamma_{YM}^n[V^{++}]&=&
\frac{i^{n+3}}{n(2\pi )^{4n}} {\rm Tr}\prod^n_{t=1}
\int
\frac{{\rm d}^4p_t{\rm d}^4\theta_t{\rm d}u_t{\rm d}\vv_t}
{(p_t^2-m^2)}
\,[{\cal D}^+(\theta_t,\vv_t)]^2\delta^4(\theta_t-\theta_{t+1})\nn
&\times&\frac{V^{++}_\cc(t)}{(u_t^+\vv_{t-1}^{+})(\vv_{t-1}^{+}\vv_t^{+})(\vv_t^{+}u_t^+)}
\int{\rm d}^4{\bar{\theta}}_1{\rm d}^4{\bar{\theta}}_2
\delta^4({\bar\theta}_1-{\bar\theta}_2)
\,[\mbox{\bf Ch}]\,
\delta^4({\bar\theta}_2-{\bar\theta}_1)
 + c.c.~, \label{Ch}
\eea
where by the symbol $[\mbox{\bf Ch}]$ we denoted 
the following chain of differential operators:
\be
[\mbox{\bf Ch}] \equiv [\bar{\cal {D}}^+(\vv_1)]^2[\bar{\cal {D}}^+(\vv_2)]^2\dots
[\bar{\cal {D}}^+(\vv_n)]^2~. \label{chain}
\ee
Reducing this chain as explained in Appendix A, 
doing the ${\bar{\theta}}_2$ integral with making use 
of the relation 
\bea
\int{\rm d}u\frac{u^+_iu^+_j}{(u^+{\bf u}^+)(u^+{\scriptstyle \cal V}^+)}=
\frac{{\bf u}^+_{(i}{\bf u}^-_{j)}}{{\scriptstyle \cal V}^+{\bf u}^+}-
\frac{{\scriptstyle \cal V}^+_{(i}{\scriptstyle 
\cal V}^-_{j)}}{{\scriptstyle \cal V}^+{\bf u}^+}~, 
\eea
and, finally, performing the integration over the harmonics $\vv_1, \vv_2, 
\dots \vv_n\;,$ 
we reduce (\ref{start}) to 
\bea
\!\!\!\!\!\Gamma_{YM}^n[V^{++}] &=&
\frac{i^{2n-1}W^{n-2}_0}{4^n(2\pi )^{4n}n} {\rm Tr}\prod^n_{t=1}
\int
\frac{{\rm d}^4p_t{\rm d}^4\theta_t{\rm d}u_t
{\rm d}^4{\bar{\theta}}}
{(p_t^2-m^2)}
\,\frac{ V_\cc^{++}(t)}
{(u_t^+u_{t+1}^+)}\nn
&\times&\!\!\!\!\Big\{[{\cal D}^+(\theta_t,u_t){\cal D}^-(\theta_t,u_t)]
-[{\cal D}^+(\theta_{t+1},u_{t+1}){\cal D}^-(\theta_{t+1},u_{t+1})]\Big\}
\delta^4(\theta_t-\theta_{t+1}) + c.c.
\eea
Thanks to the identity
\bea
\label{ident}
{\cal D}^-(u_n)(u^+_nu^+_p) ={\cal D}^+(u_p) + 
{\cal D}^+(u_n)(u_n^-u_p^+)\,,
\eea
the previous expression can be rewritten in the form
\bea
\label{big}
\Gamma_{YM}^n[V^{++}]=
\frac{i^{2n-1}}{n(2\pi )^{4n}} {\rm Tr}\int{\rm d}^4{\bar{\theta}}
\prod^n_{t=1}\int\;\frac{{\rm d}^4p_t{\rm d}^4\theta_t{\rm d}u_t}
{(p_t^2-m^2)}
\, \frac{V_\cc^{++}(t)}
{(u_t^+u_{t+1}^+)^2} \;[A_n+B_n]\,W^{n-2}_0 \,+ c.c.\,,
\eea
where
\bea
A_n &=& {1\over 4^n}\prod^n_{s=1}
 2({\cal D}_s^+{\cal D}_{s+1}^+)
\delta^4(\theta_s-\theta_{s+1}) \nn
B_n &=&\prod^n_{s=1}
[({\cal D}^+_s)^2(u_s^-u_{s+1}^+) + ({\cal D}^+_{s+1})^2(u_{s+1}^-u_s^+)]
\delta^4(\theta_s-\theta_{s+1})~. \nonumber 
\eea

Let us first consider the term proportional to $A_n$ in ~(\ref{big}).
Rearranging the derivatives in it and integrating over $\theta_3 ,\dots
,\theta_n$, we extract the following terms leading to the holomorphic 
and anti-holomorphic contributions
\bea
&&\frac{2 i^{2n+1}{W}_0^{n-2}}{(2\pi)^{4n}}
 {\rm Tr}\int{\rm d}^4\theta_1{\rm d}^4
\theta_2{\rm d}^4\bar{\theta}
\prod^n_{t=1}\int\frac{{\rm d}^4p_t{\rm d}u_t}
{(p^2_t-m^2)}
\,\prod^n_{s=3}\frac{({\cal D}^+_{s-1})^2V^{++}_\cc(\theta_1,u_s)}
{(u_{s-1}^+u_s^+)^2}\nn
&&\times
\frac{{V}^{++}_\cc(\theta_1,u_1){V}^{++}_\cc(\theta_2,u_2)}
{(u_1^+u_2^+)^2}
\,\delta^4(\theta_1-\theta_2)({\cal D}^+_n)^2({\cal D}_1^+)^2
\delta^4(\theta_1-\theta_2)
 + c.c.~. \label{nu}
\eea
The superfield $V^{++}_\cc$ is invariant under the central charge, 
hence ${\cal D}^+V_\cc=D^+V_\cc$.
Now, using the identity (\ref{ident}) and 
analyticity of $V^{++}_\cc$, we obtain
\bea
\label{D_D}
({\cal D}^+_{m-1})^2V_\cc^{++}(u_m)=(D^-_m)^2V_\cc^{++}(u_m)(u^+_{m-1}u^+_m)^2\, .
\eea
Using eqs. (\ref{D_D}) and (\ref{W_Vc}), we find that (\ref{nu})
in the local limit is reduced to 
\bea
\label{1}
&&2\frac{i}{n}\int\frac{{\rm d}^4p}
{(2\pi)^{4n}(p^2-W_0{\bar{W}}_0)^n}
\int{\rm d}^4x{\rm d}^4\bar\theta\,{W}_0^{n-2}{\bar{W}}_\cc^n + c.c.
\eea
(from here on, we denote by $W_0$, $W_\cc, \ldots $ 
the coefficients of $T_3$ in the corresponding $U(1)$ 
subalgebra-valued quantities).

The second term in ~(\ref{big}) can be handled similarly. Using the identity 
$u_{2j}^+u_2^{i-}=u_{2j}^-u_2^{+i}-\delta^i_j$ and integrating by parts,
we reduce it to the form in which it contains no explicit harmonics.
In the local limit it yields
\bea
\label{2}
&&\frac{i^{2n+3}}{2^{n-2}n}\int\frac{{\rm d}^4p}
{(2\pi)^{4n}(p^2-W_0{\bar{W}}_0)^n}
\int{\rm d}^4x{\rm d}^4\bar\theta\,{W}_0^{n-2}{\bar{W}}_\cc^n + c.c.\,.
\eea

Putting together ~(\ref{1}) and ~(\ref{2}) results in 
\bea
\label{ym_n}
\Gamma_{YM}^n[V^{++}] &=&
2\,\frac{i}{n}\,\int\frac{{\rm d}^4p}
{(2\pi)^{4n}(p^2-W_0{\bar{W}}_0)^n}
\int{\rm d}^4x{\rm d}^4\bar\theta\,{W}_0^{n-2}{\bar{W}}_\cc^n \nonumber \\
&+& \frac{i(-1)^{n+1}}{2^{n-2}n}\int\frac{{\rm d}^4p}
{(2\pi)^{4n}(p^2-W_0{\bar{W}}_0)^n}
\int{\rm d}^4x{\rm d}^4\bar\theta\,{W}_0^{n-2}{\bar{W}}_\cc^n + c.c.\,.
\eea

The ghost contribution is given by the supergraphs on Fig. 3.
The sum of holomorphic and anti-holomorphic contributions 
from $\Gamma_{gh}^n[V^{++}]$
coincides, up to a coefficient, with the second term in eq. (\ref{big}) 
(that is proportional to $B_n$). In the local limit it is 
\bea
\label{gh_n}
&&\Gamma_{gh}^n[V^{++}]=\frac{i(-1)^n}{2^{n-2}n}\int\frac{{\rm d}^4p}
{(2\pi)^{4n}(p^2-W_0{\bar{W}}_0)^n}
\int{\rm d}^4x{\rm d}^4\bar\theta\,{W}_0^{n-2}{\bar{W}}_\cc^n + c.c.\,.
\eea

Putting together~(\ref{ym_n}) and~(\ref{gh_n}), we get the full  
$n$-th order contribution from the $N=2$ SYM sector
\bea
\Gamma_{YM}^n[V^{++}]+\Gamma_{gh}^n[V^{++}] =
2\,\frac{i}{n}\,\int\frac{{\rm d}^4p}
{(2\pi)^{4n}(p^2-W_0{\bar{W}}_0)^n}
\int{\rm d}^4x{\rm d}^4\bar\theta\,{W}_0^{n-2}{\bar{W}}_\cc^n
 + c.c.\,. \label{totaln}
\eea
The total low-energy holomorphic effective action is obtained 
by summing over $n$ and making the appropriate 
renormalization in the second-order term. 
The final answer is as follows
\bea
&&\Gamma_{YM}[V^{++}]+\Gamma_{gh}[V^{++}]=\frac{1}{32\pi^2}\,\int{\it d}^4x
{\rm d}^4\theta\,
W_\ccc^2\ln\frac{W_\ccc^2}{M^2}+c.c.\;,\label{ym_t} \\
&&W_\ccc=W_0+W_\cc\;, \nonumber
\eea
with $M$ being normalization scale. It coincides 
(up to a difference in the conventions) 
with Seiberg's action \cite{seib} obtained by 
integrating $U(1)_R$ anomaly in the pure $N=2$ SYM theory. 
The same arguments as in ref. \cite{bbik} show that 
the holomorphic contribution (\ref{ym_t}) is entirely due to  
non-zero central charges and is vanishing in the limit 
$W_0 = \bar W_0 = 0$ (this does not apply to the divergent 
second-order correction which exists in the massless case 
as well \cite{gios1}). The same holomorphic 
contribution was found in \cite{bfm} using the background fields method. 
In our case we calculated it using the computation techniques 
manifestly covariant under $N=2$ SUSY with the central charges. 

The calculation of the hypermultiplet contribution 
follows the same lines as in the abelian case \cite{bbik}. 
Here we present only the final answer for the hypermultiplet 
in the adjoint representation. It amounts to the 
contribution of two charged massive hypermultiplets 
(two off-diagonal components of the adjoint representation 
of $SU(2)$) and differs only in sign from the total 
contribution of the gauge sector:
\bea
\label{om_t}
&&\Gamma_{HP}[V^{++}]=-\frac{1}{32\pi^2}\,\int{\it d}^4x{\rm d}^4\theta\,
W_\ccc^2\ln\frac{W_\ccc^2}{M^2}+c.c.\,.
\eea
The cancellation among the $N=2$ SYM and hypermultiplet contributions 
agrees with the absence of holomorphic corrections 
to the $N=4$ SYM effective 
action in the $N=2$ superfield notation and serves 
a good consistency check of our computation of the 
$N=2$ SYM contribution (\ref{ym_t}). Note that 
in the present case we deal with a bit different 
$N=4$ SYM action: both manifest and hidden $N=2$ SUSY 
in it possess central charges proportional to $W_0, \bar W_0$. They break 
the $U(4)_R$ symmetry of $N=4$ SUSY down to $SO(4)$ which is the product 
of the $N=2$ SUSY automorphism $SU(2)$ and the so-called Pauli-G\"ursey 
$SU(2)$ realized on the doublet indices $a$ in (\ref{hyp2}).   

Being armed with the above techniques, we can 
compute the holomorphic contribution 
to the $N=2$ SYM effective action for the case of 
arbitrary semi-simple gauge group $G$ of rank $r$. 
It is convenient to choose the standard 
Cartan-Weyl basis for its algebra $\ag$  
\bea
\label{decomp}
\ag=n_+\oplus h\oplus n_-\; ,
\eea
where $h$ is the Cartan subalgebra with
the basis $\{H_{i}\},\; i=1\dots r$, and $n_+$, $n_-$ 
are the positive and negative roots subspaces spanned 
by the generators $E_{(+\lambda)}$, $E_{(-\lambda})$ ($\pm \lambda$ 
are the positive (negative) roots). For our purposes we need 
only the commutation relations involving the Cartan generators 
\bea
&& [H_{i},\;H_{k}]=0\;,\quad
[H_{i},\;E_{(\pm \lambda)}]= \pm \lambda^{(i)}E_{(\pm\lambda)}\;, \quad
[E_{(+\lambda)},\;E_{(-\lambda)}]= \lambda^{(i)}H_i\;,  \label{alg} \\
&& \mbox{Tr}(H_iH_k) = \delta_{ik}~, \quad \mbox{Tr}(E_{(+\lambda)}
E_{(-\beta)}) = 
\delta_{\lambda\,\beta}~. \label{algtr} 
\eea 
Next we notice that all the vertices we use in our calculations 
can be uniformly written as follows 
\be
\sim \mbox{Tr} \left([M,\;K] V^{++}_\cc \right) = 
\mbox{Tr} \left([V^{++}_\cc, \;M] K  \right)  = 
\mbox{Tr} \left(M [K,\; V^{++}_\cc] \right)~, \label{vert}
\ee
where
$$
\quad V^{++}_\cc = 
\sum_i \tilde{V}^{++}_{i} H_i~,  
$$
and 
$$
M = \sum_\lambda M_{(-\lambda)} E_{(+\lambda)}~, \qquad 
K = \sum_\beta K_{(+\beta)} E_{(-\beta)} 
$$
stand for the positive (negative) roots parts either of the 
Faddeev-Popov ghost superfields, or the hypermultiplet in 
the adjoint representation, or the massive $N=2$ SYM potentials 
$\hat V^{++}(1), \hat V^{++}(2)$. Defining 
\be
V^{++}_{(\lambda)}E_{(+\lambda)} \equiv [V^{++}_\cc,\; E_{(+\lambda)}]
\ee
and using (\ref{algtr}), we represent these vertices as 
\be
\sim \sum_\lambda M_{(-\lambda)}\,V^{++}_{(\lambda)}\, K_{(+\lambda)}~.
\ee
Then one should take into account that only the propagators 
$<M_{(-\lambda)}(1) \vert K_{(+\beta)}(2)>$ appear as the internal 
lines in the one-loop diagrams 
we are interested in, and that these propagators are diagonal in 
the indices $\lambda, \beta$ (because they include only 
the unity matrix and the diagonal matrices 
of the Cartan generators). 
With this in mind, it is easy to find that 
the full holomorphic and anti-holomorphic correction 
from the SYM and ghost superfields is given by the sum 
over the positive roots, like in refs. \cite{roots}
\bea
\Gamma_{YM}+\Gamma_{gh}=
\frac{1}{32\pi^2}\,\sum_{\lambda \in n^+\;}\int{\it d}^4x
{\rm d}^4\theta\,
W^2_{(\lambda)}\ln\frac{W^2_{(\lambda)}}{M^2}+c.c.\;, \qquad
W_{(\lambda)}E_{(+\lambda)} \equiv [W_{\cc},\; 
E_{(+\lambda)}]~. \label{end}
\eea
As before, the hypermultiplet contribution differs only by 
sign from (\ref{end}) (once again, we specialize to the adjoint 
representation). Thus, like in the $SU(2)$ case, we have for 
the $N=4$ SYM holomorphic effective action
\bea
\Gamma_{N=4}^{hol}[V^{++}]=\Gamma_{HP}^{hol}[V^{++}]+
\Gamma_{YM}^{hol}[V^{++}]+\Gamma_{gh}^{hol}[V^{++}]=0\;.
\eea

Finally, let us make two comments.

The first one concerns the gauge invariance of our approach. 
Our computations are clearly covariant under the Cartan subgroup 
$C$ of the full gauge group $G$. What concerns non-abelian 
gauge $G/C$ transformations, 
their sole effect seems to be rotations 
of the considered sector of the full $N=2$ SYM effective action into other 
sectors, with some numbers of off-diagonal components of the non-abelian 
strength as external legs. As for the internal lines, 
the above diagrams exhaust all possibilities in the sector 
we are considering, provided that one is interested only 
in the holomorphic contributions and 
keeps in mind the property discussed in the end of Sect.~3. 
In Appendix B we also argue that our results cannot depend 
on the choice of the $\alpha$ gauge in the massive 
$\hat{V}^{++}$ propagator. Our reasoning is based on 
an explicit calculation. An implicit argument why the full 
$N=2$ SYM contribution should be $\alpha$-independent is 
the absence of holomorphic corrections to the $N=4$ SYM action. 
Indeed, this contribution should be cancelled by that 
from the $q$-hypermultiplet sector, but the latter contains 
no any $\alpha$-dependence.

Secondly, we wish to note that 
in the one-loop approximation all possible diagrams with massless 
gauge superfields 
outside necessarily contain only massive $V^{++}$ lines inside. This 
follows from the conservation of the Cartan $U(1)$ charges in the 
diagram and the property that among the elementary $N=2$ SYM vertices 
there exist no those containing only Cartan $V^{++}_\cc$ 
components of $\tilde{V}^{++}$: any vertex should have 
at least two massive $\hat{V}^{++}$. To be 
convinced 
of this, it is enough to observe that the nominator of a generic term 
in the 
$N=2$ SYM self-interaction (\ref{selfN2}) can be rewritten 
through the successive commutators as
$$
\sim \mbox{Tr}\{[(\tilde{V}^{++}_1)_\tau\;,[(\tilde{V}_2^{++})_\tau\;,....,
[(V^{++}_{n-2})_\tau\;, (V^{++}_{n-1})_\tau]...]](V^{++}_{n})_\tau \}\;.
$$
So, all possible one-loop diagrams with the $V^{++}_\cc$ 
external legs contain 
inside only the massive $\hat{V}^{++}$ lines, 
and we can apply to them the general argument 
adduced in the end of Sect.~3.

\setcounter{equation}{0}          
\section{Conclusions}
In this paper we worked out the basic elements of the quantum HSS techniques for 
non-abelian $N=2$ SYM theory in the Coulomb branch, with $N=2$ SUSY modified 
by the Cartan central charges which give rise to the splitting of 
$N=2$ SYM prepotential $V^{++}$ into the massless Cartan and massive 
charged $G/C$ projections. Using this approach, we calculated 
the holomorphic contribution of the massive $N=2$ SYM superfields (and ghosts) 
into the effective action of the massless projection. Up to a numerical 
coefficient, it has the same form as the contribution of one $U(1)$ 
$q^+$ hypermultiplet computed earlier in \cite{bbik} 
with making use of similar techniques. Like the latter, it disappears in the 
limit of zero central charge, thus confirming the general property 
that all holomorphic 
perturbative one-loop contributions to the $U(1)$ gauge superfields sector of the 
quantum effective action of $N=2$ SYM in the Coulomb branch are 
the entire effect of non-zero central charge in 
$N=2$ superalgebra (similarly to the analytic 
corrections to the $q^+$ effective action \cite{ikz}). We found 
the explicit 
cancellation of the $N=2$ SYM contribution and that of 
the matter hypermultiplet in the 
adjoint representation of the gauge group, thus having 
checked the absence of the 
one-loop holomorphic corrections to $N=4$ SYM action. 
The main novelty of our 
approach is the preservation of manifest $N=2$ SUSY with the 
Cartan central charges 
at all steps of quantum calculations in non-abelian $N=2$ SYM theories.         

It would be interesting to apply the same straightforward 
quantum HSS techniques to computing non-holomorphic 
contributions to the Coulomb branch effective actions 
of $N=2$ and $N=4$ SYM theories, both in the gauge fields 
and hypermultiplet sectors. 

\section*{Acknowledgments}
It is a pleasure for us to thank Joseph Buchbinder for useful remarks and 
stimulating discussions. A support from the grants RFBR-99-02-18417, 
INTAS-96-0308 and INTAS-96-0538 is cordially acknowledged. 

\section*{Appendix A}

\setcounter{equation}{0}
\def\theequation{A.\arabic{equation}}

Here we show how to simplify the chain operator $[\mbox{\bf Ch}] =
[\bar{\cal {D}}^+(u_1)]^2[\bar{\cal {D}}^+(u_2)]^2\dots \break
[\bar{\cal {D}}^+(u_n)]^2$ appearing in eq. (\ref{Ch}). 
We follow the method proposed in \cite{bbik}.

Let us write this chain of derivatives in the following form

\bea
\label{31.5}
\prod^n_{t=1}[{\cal D}^+(u_t)]^2=
\displaystyle\frac{1}{4^n}
\prod^{n}_{t=1} ({\cal D}_t^+ {\cal D}_t^+)
=\displaystyle\frac{(-1)^{n-1}}{4^n}
\Big\{\prod^n_{t=1}(u^+_t u^+_{t+1})
({\cal D}_t^{+} {\cal D}_{t+1}^-)\Big\}
\{{\cal D}_n^{+} {\cal D}_{n}^+\}\;.
\eea
Here all the central basis spinor derivatives are evaluated 
at the same $\theta$-point.

As the next step, we expand the derivative
${\cal D}_{n-1}^{+ \alpha}$ in (\ref{31.5})
over the $n$-th set of harmonics. The result reads
\bea
\label{32}
\prod^n_{t=1}[{\cal D}^+(u_t)]^2=&&
\displaystyle\frac{(-1)^{n-1}}{4^n}
\Big\{\prod^{n-1}_{t=1}(u^+_t u^+_{t+1})\prod^{n-2}_{t=1}
({\cal D}_t^{+} {\cal D}_{t+1}^-)  \Big\} \nn
&&\times\Big\{({\cal D}_{n}^{-} {\cal D}_{n}^-)
({\cal D}_n^{+} {\cal D}_{n}^+)(u^+_n u^+_
{n-1})
 -
({\cal D}_{n}^{+} {\cal D}_{n}^-)
({\cal D}_n^{+} {\cal D}_{n}^+)(u^-_n u^+_{n-1}) \Big\}\;.
\eea
Further, in the first term we expand ${\cal D}^-_{n-1}$ again in terms
of the $n$-th harmonics (we suppress spinor indices).
Only the ${\cal D}^+_{n}$ projection
survives,
then we anticommute it with ${\cal D}^-_{n}$ using the
algebra (\ref{derev}). Since $({\cal D}^+)^3=0$, 
only the commutator remains.
The situation is simpler with the second term. We just anticommute
${\cal D}^{+}_{n}$ with ${\cal D}^-_{n}$  and
then expand ${\cal D}^-_{n-1}$ over the $n$-th harmonics.
We arrive at
\bea
\label{33}
\prod^n_{t=1}[{\cal D}^+(u_t)]^2=&&
\displaystyle\frac{(-1)^{n-1}}{4^n} 4{\rm i}\bar{W_0}
\Big\{\prod^{n-1}_{t=1}(u^+_t u^+_{t+1})\prod^{n-3}_{t=1}
({\cal D}_t^{+} {\cal D}_{t+1}^-)  \Big\} \nn
&&\times\Big\{({\cal D}^{+}_{n-2} {\cal D}^-_{n})
({\cal D}^{+}_{n} {\cal D}^-_{n})
[ (u^+_n u^+_{n-1})(u^-_n u^-_{n-1})- (u^+_n u^-_{n-1})
(u^-_n u^+_{n-1})] \Big\}\;.
\eea
Here the harmonic expression in the
square brackets equals 1 due to the harmonics completeness relation, 
and we finally get
\bea
\label{34}
\prod^n_{t=1}[{\cal D}^+(u_t)]^2=&&
\displaystyle\frac{(-1)^{n-1}}{4^n} 4{\rm t}\bar W_0
\Big\{\prod^{n-1}_{t=1}(u^+_t u^+_{t+1})\prod^{n-3}_{t=1}
({\cal D}_t^{+} {\cal D}_{t+1}^-)  \Big\}
({\cal D}^{+}_{n-2} {\cal D}^-_{n})
({\cal D}^{+}_{n} {\cal D}^-_{n})\;.
\eea

Comparing (\ref{34}) with (\ref{31.5}), we observe that 
the former can be formally
obtained from the latter by replacing the block
${\cal D}^-_{ n-1 \alpha}{\cal D}^{+ \beta}_{n-1}$
by $4{\rm i}\delta^{\beta}_{\alpha}$.  This observation allows the
process to go on. The chain $[{\cal D}^+(u_1)]^2 [{\cal
D}^+(u_2)]^2 \dots [{\cal D}^+(u_n)]^2$ is finally
reduced to
$$
\displaystyle\frac{-(-{\rm t})^{n-2}}{4^2} \bar{W_0}^{n-2}
\Big\{\prod^{n-1}_{t=1}(u^+_t u^+_{t+1})\Big\}
({\cal D}^{+}_1
{\cal D}^-_{n})({\cal D}_n^+{\cal D}^+_n)\;.
$$
Recalling that the chain operator in eq. (\ref{Ch}) is sandwiched between two chiral 
harmonic $\delta$-function and using (\ref{deltaderev}), the latter 
expression can be finally cast in the form 
\bea
-(-i)^{n-2}\bar{W_0}^{n-2}\Big\{\prod^n_{t=1}(u^+_t u^+_{t+1})\Big\}\;.
\eea

\section*{Appendix B}

\setcounter{equation}{0}
\def\theequation{B.\arabic{equation}}

In this Appendix we argue that our computations do not depend on the 
choice of $\alpha$-gauge for the massive $\hat{V}^{++}$ propagators.
To this end, let us start with the full $\alpha$-depended 
propagator which is a solution of the equation (\ref{gren}).
\bea
<({\hat{V}}^{++}_1)_\tau\mid({\hat{V}}^{++}_2)_\tau>_{(\alpha)}=
\dis\frac{2}{\bbx_1}\,({\cal D}^+_1)^4\,
\delta^{12}(X_1-X_2)\,\delta^{(-2,2)}(u_1,u_2)-
2\,\frac{1+\alpha}{\bbx_1}\,\Pi {}^{(2,2)}(1\mid 2)\;.
\eea 
Here $\Pi$ is the projection operator
\bea
\Pi {}^{(2,2)}(1\mid 2)=
\frac{({\cal D}^+_1)^4({\cal D}^+_2)^4}{\bbx_1}\,\frac{\delta^{12}(X_1-X_2)}
{(u_1^+u_2^+)^2} \;.
\eea
Now we substitute this propagator into $\Gamma_{YM}^n$ (\ref{ym}) and 
decompose the latter in series in $\alpha^\prime=\alpha+1$ at the point 
$\alpha^\prime = 0\; (\alpha = -1)$
\bea
\label{apb}
\Gamma_{YM}^n(\alpha^\prime)=
\Gamma_{YM}^n +\Delta^n\;\delta \alpha^\prime + \dots \; .
\eea
The idea is to show that, under the infinitesimal shift of
 $\alpha^\prime$, the holomorphic (anti-holomorphic) part of
$\Gamma_{YM}^n$ does not change, i.e. $\Delta^n = 0$ in the local 
limit. Then, assuming $\Gamma_{YM}^n (\alpha^\prime)$ 
to be continuous in $\alpha^\prime$, we can state that 
its (anti)holomorphic local part does not depend on 
$\alpha^\prime$ at all.  

We find 
\bea
\Delta^n\;\sim&&
\frac{(-i)^{n+1}}{n}\prod^n_{t=1}
\int
{\rm d}^4x_t{\rm d}^8\theta_t{\rm d}u_t{\rm d}\omega_t{\rm d}\vv_t
\;\frac{V_\cc^{++}(t)}
{(u_t^+\om_t^{+})(\om_t^{+}\vv_t^{+})(\vv_t^{+}u_t)}\nn
&&\times\prod^{n-1}_{t=1}\frac{1}{\Box_t+m^2}
[{\cal D}^+(\theta_t,\vv_t)]^4\delta^{8}(x_t-x_{t+1})
\delta^8(\theta_t-\theta_{t+1})\delta^{(-2,2)}(\vv_t,\om_{t+1})\nn
&&\times\frac{[{\cal D}^+(\theta_n,\vv_t)]^4[{\cal D}^+(\theta_1,\omega_1)]^4}
{(\Box_t+m^2)^2(u_1^+u_n^+)^2}
\delta^{8}(x_t-x_{t+1})
\delta^8(\theta_t-\theta_{t+1})\;.
\eea
Then, following the calculations (\ref{ym})-(\ref{big}), 
we get the expression which looks very similar to (\ref{big}), 
except for the presence of an extra pair of spinor derivatives 
$({\cal D}^+_l{\cal D}^+_k)$ in it:
\bea
\Delta^n\;\;\sim&& 
\int{\rm d}^4{\bar{\theta}}
\prod^n_{t=1}\int
\frac{{\rm d}^4p_t{\rm d}^4\theta_t{\rm d}u_t}
{(p_t^2-m^2)}
\times{V_\cc^{++}(t)}\nn
&&\times\frac{f(u_1,\dots ,u_n)W^{n-1}_0}{(p_n^2-m^2)}\times
({\cal D}^+_l{\cal D}^+_k) [A_n+B_n]
\qquad + \mbox{c.c.}\;. \label{Delta} 
\eea
But (\ref{big}) already contains the critical number 
of spinor derivatives and the presence of two extra 
ones simply means that the holomorphic 
(anti-holomorphic) contribution from (\ref{Delta}) vanishes 
\bea
\Delta^n_{hol} =0
\eea
(due to the properties $({\cal D}^+_\alpha)^3=0\;$ and 
$\;{\cal D}^+_\alpha(u)V^{++}_\cc(u)=0)$).

We have also explicitly checked the vanishing of 
the holomorphic contributons in the second order in $\delta \alpha'$ 
for the diagrams with two and three external legs.

\end{document}